\documentclass[10pt, a4paper]{article}
\usepackage{hyperref}
\usepackage{times}
\usepackage{latexsym}
\usepackage{comment}

\usepackage[T1]{fontenc}

\usepackage[utf8]{inputenc}

\usepackage{microtype}

\usepackage{inconsolata}
\usepackage{booktabs}
\usepackage{graphicx}
\usepackage{amsmath}
\usepackage{lipsum}
\usepackage{xstring}
\usepackage{cleveref}
\newcommand{\datasetraw}[0]{$\mathcal{D}_\text{raw}$}
\newcommand{\datasetmanual}[0]{$\mathcal{D}_\text{manual}$}
\newcommand{\datasetrawdp}[0]{$\text{P}_\texttt{DP}$}
\newcommand{\datasetrawNER}[0]{$\text{P}_\texttt{NER}$}
\newcommand{\datasetLLM}[0]{$\text{P}_\texttt{LLM}$}
\newcommand{\datasetNERdp}[0]{$\text{P}_{\texttt{NER}\rightarrow\texttt{DP}}$}
\newcommand{\datasetLLMdp}[0]{$\text{P}_{\texttt{LLM}\rightarrow\texttt{DP}}$}

\usepackage[final]{lrec2026} 

\title{Differentially Private De-identification of Dutch Clinical Notes: A Comparative Evaluation}

\name{\parbox{\textwidth}{\centering
Michele Miranda$^{1}$$^{2}$$^{*}$\thanks{$^{*}$These authors contributed equally to this work.}, Xinlan Yan$^{3}$$^{4}$$^{*}$\footnotemark[1], Nishant Mishra$^{3}$$^{4}$,\
Rachel Murphy$^{3}$$^{4}$, \\ Ameen Abu\mbox{-}Hanna$^{3}$$^{4}$, Sébastien Bratières$^{2}$, Iacer Calixto$^{3}$$^{4}$
}}

\address{$^{1}$Sapienza University of Rome, $^{2}$Translated, $^{3}$Amsterdam UMC, $^{4}$University of Amsterdam \\
         miranda@di.uniroma1.it, sebastien@translated.com \\
         \{x.yan, n.mishra, r.m.murphy, a.abu-hanna, i.coimbra\}@amsterdamumc.nl\\
         }

\abstract{
Protecting patient privacy in clinical narratives is essential for enabling secondary use of healthcare data under regulations such as GDPR and HIPAA.
While manual de-identification remains the gold standard, it is costly and slow, motivating the need for automated methods that combine privacy guarantees with high utility.
Historically, most automated text de-identification pipelines employed named entity recognition (NER) to identify protected entities for redaction.
Although methods based on differential privacy (DP) provide formal privacy guarantees, more recently also large language models (LLMs) are increasingly used for text de-identification in the clinical domain.
In this work, we present the first comparative study of DP, NER, and LLMs for \textit{Dutch} clinical text de-identification.
We investigate these methods separately as well as hybrid strategies that apply NER or LLM preprocessing prior to DP, and assess performance in terms of privacy leakage and extrinsic evaluation (entity and relation classification).
We show that DP mechanisms alone degrade utility substantially, but combining them with linguistic preprocessing, especially LLM-based redaction, significantly improves the privacy–utility trade-off.
\footnote{We will release all code to reproduce our experiments upon acceptance.}
 \\ \newline  }

\begin{document}

\maketitleabstract

\section{Introduction}\label{sec:intro}
Ensuring privacy in clinical texts is critical to enable data sharing for healthcare research~\citep{conduah2025data}.
Privacy regulations like GDPR~\citep{gdpr} and HIPAA~\citep{hipaa} require the redaction or \textit{de-identification} of all personally identifiable information (PII) to protect patient privacy.

Methods for PII de-identification based on named entity recognition (NER) have been extensively used for English and other languages~\cite{grouin-etal-2015-possible,Dernoncourt2017-aa,Bourdois_Avalos_Chenais_Thiessard_Revel_Gil-Jardine_Lagarde_2021,tchouka2022easy,wang2022-uu}.
More recently, large language models (LLMs) have been applied to PII de-identification and have shown strong performance~\cite{liu2023deidgptzeroshotmedicaltext}.
NER- and LLM-based methods, despite their performance, do not provide any formal privacy guarantees.
Differential privacy~\citep[DP;][]{10.1007/11787006_1,10.1561/0400000042}, on the other hand, offers a principled mechanism with formal guarantees against privacy leakage when sharing a \textit{privatized} dataset~\citep{Chatzikokolakis2013BroadeningTS, tong2025inferdptprivacypreservinginferenceblackbox}.

The tension in the existing literature lies in the fact that DP-based methods for text de-identification provide formal privacy guarantees but can severely impact utility \cite{Yu2021DifferentiallyPF,Yue2022SyntheticTG,tchouka2022easy,tchouka2022identification}, whereas LLM-based methods become increasingly strong but do not provide any privacy guarantees~\cite{pissarra-etal-2024-unlocking,yang-etal-2025-robust}.
These LLM-based methods remain limited as they redact only detected entities rather than full text---offering weaker privacy preservation than DP---, overlook instruction-tuned LLMs, demand computational resources often unavailable in privacy-critical settings, and are developed solely for English and are thus untested in other languages.
\begin{figure}[t!]
     \centering
     \includegraphics[width=170pt]{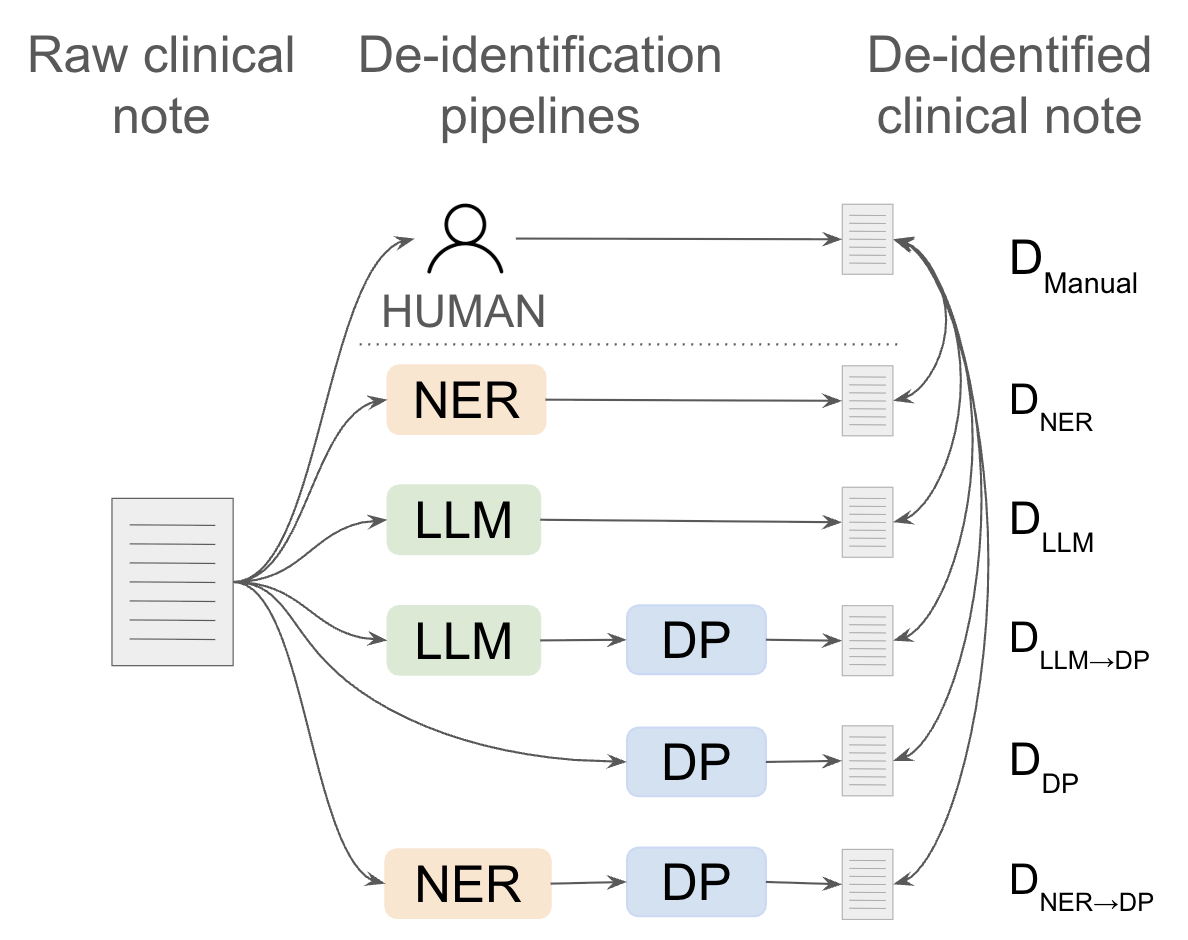}
     \caption{Overview of our comparative analysis. A raw document \datasetraw{} is de-identified using 5 different pipelines, which are evaluated against a manually de-identified version of the same document \datasetmanual{}. We use a range of open-source and proprietary LLMs that vary in architecture and size in our experiments.}
     \label{fig:overview}
\end{figure}

In this study, we compare different PII de-identification methods using state-of-the-art NER~\citep{zaratiana-etal-2024-gliner}, LLMs~\citep{liu2023deidgptzeroshotmedicaltext}, and synthetic data generation methods with DP guarantees~\citep{Chatzikokolakis2013BroadeningTS,tong2025inferdptprivacypreservinginferenceblackbox} for Dutch clinical notes. See Figure~\ref{fig:overview} for an overview of our comparisions. Unlike prior works on English or, more recently, French clinical text~\citep{tchouka2022a,tchouka2022b}, our work is \textbf{the first study to investigate DP-based text anonymization in Dutch real-world hospital clinical notes}, addressing a critical gap in the de-identification research.

We evaluate the quality of the de-identified clinical notes \textit{intrinsically} by quantifying the remaining residual PII after de-identification, and \textit{extrinsically} by measuring the performance of prediction models trained using the generated de-identified notes for two downstream tasks: entity classification (drugs and disorders) and relation classification (adverse drug events).
Finally, we also combine a strong de-identification method (e.g. NER- or LLM-based) as a preprocessing step with DP de-identification.
We hypothesise that by doing this we can considerably reduce the DP noise necessary to achieve the same privacy guarantee, possibly leading to a better privacy-utility trade-off.

\typeout{=== RELATED WORKS INCLUDED ===}
\section{Related Works}\label{sec:related}

De-identification of clinical text is essential to allow secondary use of healthcare data while protecting patient privacy under privacy regulations such as HIPAA \citep{hipaa} and GDPR \citep{gdpr}. Traditional manual de-identification methods are labour-intensive, costly, and prone to inconsistencies. This section synthesizes key methodological advancements in automated PII de-identification.

\paragraph{Differential Privacy Approaches}

Differential Privacy \citep{10.1007/11787006_1} offers formal privacy guarantees by ensuring that the outputs of algorithms are statistically indistinguishable across neighboring datasets.
In other words, the key idea is that two models ($\mathcal{M}_1$, $\mathcal{M}_2$) trained on two datasets ($\mathcal{D}_1$, $\mathcal{D}_2$) that differ only by one entry, i.e., one patient record, should be indistinguishable in terms of their predicted outcomes.

Metric DP extends the traditional DP framework by introducing a metric space that quantifies the similarity between data points \citep{Chatzikokolakis2013BroadeningTS}. In the context of text data, Metric DP leverages semantic similarity measures to guide the privacy mechanism. \citet{Feyisetan2019LeveragingHR} applied Metric DP to word embeddings, enabling controlled perturbations that account for the semantic relationships between words. This results in text transformations that better preserve the utility of the original data while providing privacy guarantees.

Another promising approach is RANTEXT \citep{tong2025inferdptprivacypreservinginferenceblackbox}, a token-level DP framework designed for privacy-preserving text generation. RANTEXT dynamically constructs randomized adjacency lists for context-aware token substitution and has demonstrated strong resistance to membership inference attacks, outperforming earlier models like SANTEXT+ and CUSTEXT+. In empirical evaluations, it achieved up to 98\% F1 scores for semantic preservation while maintaining privacy guarantees with $\epsilon$ values below 2. Nonetheless, DP-based methods often struggle with high-dimensional clinical narratives, where stricter privacy constraints can significantly degrade data utility \citep{tong2025inferdptprivacypreservinginferenceblackbox}.

\paragraph{LLM-Based De-identification}

Large language models can be applied to text PII de-identification in different ways.
DeID-GPT \citep{liu2023deidgptzeroshotmedicaltext}, built on GPT-4, asks the model to directly identify and redact sensitive information and integrates HIPAA identifier categories directly into its prompts, achieving 99.25\% precision and 89.73\% F1 in zero-shot redaction tasks.
This can be considered very good performance, especially given the simplicity of the approach.

We note that it is unlikely that organisations would want to use GPT-4 for this task, 
since using an online model would mean sending sensitive data off-site.
Often, this may be even forbidden under privacy regulations.
In this work, we use the OpenAI API via a cloud-based service provider in a way that preserves privacy in agreement with privacy regulations.
We understand this is not always possible for every hospital, and for that reason we include experiments with both proprietary and open-weight LLMs.  

\section{Methods and Experiments}\label{sec:methods}

Below we detail the dataset (\S\ref{sec:method-dataset}),
the de-identification modules (\S\ref{sec:method-deid}) used in our de-identification pipelines that generate privacy-preserving clinical notes (\S\ref{sec:method-pipelines}; Fig.~\ref{fig:overview}),
and the intrinsic (\S\ref{sec:intrinsic_evaluation}) and extrinsic evaluation on the entity and relation classification tasks (\S\ref{sec:extrinsic_evaluation}).

\subsection{Dataset}
\label{sec:method-dataset}
We use the Dutch ADE dataset~\cite{murphy2025creation}, a benchmark corpus containing 102 clinical notes from intensive care unit (ICU) patients annotated with entity-level labels---\textit{drug} and \textit{disorder} mentions---and relation labels between \textit{drug}-\textit{disorder} pairs.
The two relation labels annotated are \textit{adverse drug event} or \textit{ADE} (when the drug caused the disorder as an adverse event) and \textit{prescribing indication} (when the drug was prescribed to treat the disorder).
We refer to the raw corpus as \datasetraw{} and to its manually de-identified version as \datasetmanual{}.
\datasetraw{} consists of 107,110 words, 0.83\% of which are identified as PII. The PII has been redacted and replaced by 17 different placeholders, each indicating one type of PII in the data. 
There are thus 17 types of PII annotated in \datasetmanual{}. Names of the placeholders and their meaning are listed in Table~\ref{tab:placeholders}. 
For further details on the corpus, we refer the reader to \citet{murphy2025creation}.

\begin{table}[t]
\centering
\resizebox{\columnwidth}{!}{%
\begin{tabular}{ll}
\toprule
\textbf{Placeholder} & \textbf{Meaning} \\ 
\midrule
<AFDELING> & Department \\ 
<APOTHEEK> & Pharmacy \\ 
<ARTS> & Doctor \\ 
<EHR> & Electronic Health Record System \\ 
<FEESTDAG> & Holiday \\ 
<GEBOORTEDATUM> & Date of Birth \\ 
<NAAM> & Name \\ 
<RARE\_DISEASE> & Rare Disease \\ 
<RARE\_DISEASE\_TREATMENT> & Rare Disease Treatment \\ 
<REVALIDATIECENTRUM> & Rehabilitation Center \\ 
<SEIN> & Signal \\ 
<STAD> & City \\ 
<TELNR> & Telephone Number \\ 
<TRIAL-ID> & ID of Clinical Trial \\ 
<ZIEKENBOEG> & Sickbay \\ 
<ZIEKENHUIS> & Hospital \\ 
<ZKH> & Abbreviation for Hospital \\ \bottomrule
\end{tabular}}
\caption{List of placeholders and their meanings.}
\label{tab:placeholders}
\end{table}

\subsection{De-identification Modules}
\label{sec:method-deid}

We apply NER-based, LLM-based, and two DP-based text methods to de-ideintify PII: metric differential privacy~\citep[Metric-DP;][]{Chatzikokolakis2013BroadeningTS}, and RANTEXT~\citep{tong2025inferdptprivacypreservinginferenceblackbox}.

\paragraph{NER-based de-identification}
We use the off-the-shelf pretrained multilingual NER model \texttt{gliner-multi-v2.1}\footnote{\url{https://huggingface.co/urchade/gliner_multi-v2.1}}, without any fine-tuning on our clinical data. We provide it with our target entity types (see Table~\ref{tab:placeholders}, e.g., \texttt{<NAAM>}, \texttt{<ZIEKENHUIS>}) and apply it to extract labeled spans from clinical text.
Each prediction consists of text span, label, and confidence score.
To balance precision and recall, we use a confidence threshold that maximises the F-1 score on a validation set, i.e., for threshold $t \in [0,1)$ we only retain entities predicted with confidence score $c \geq t$.

\paragraph{LLM-based de-identification}
Here, we adapt the approach of \citet{liu2023deidgptzeroshotmedicaltext} and prompt an LLM in a zero-shot setting to redact predefined PII categories from Table~\ref{tab:placeholders} (e.g., \texttt{<NAAM>}, \texttt{<ZIEKENHUIS>}).
When applying LLMs directly as a de-identification pipeline (\datasetLLM{}), we experiment with several LLMs that vary in architecture, domain specialization, and size, including GPT-4o~\citep{achiam2023gpt}, DeepSeekR1 (8B and 70B)~\citep{guo2025deepseek}, LLaMA-3.1 8B~\citep{dubey2024llama}, and MedGEMMA 27B~\citep{sellergren2025medgemma}.
Please see Appendix~\ref{appendix:llm-prompt} for the prompt we use.
When using LLMs combined with DP (\datasetLLMdp{}), we use the best-performing open-source LLM in terms of privacy (Deepseek-70B).

For our experiments with GPT models, we use the OpenAI API in a privacy-compliant setting whereby no private data leaves the premises of our hospital.
Moreover, all our experiments are conducted in a privacy-compliant setting and strictly follow the data usage agreement of the Dutch ADE dataset.

\paragraph{Metric-DP}
Metric-DP introduces geometric noise in the embedding space to privatize tokens~\citep{Chatzikokolakis2013BroadeningTS}.
We use embeddings from BERTje\footnote{\url{https://huggingface.co/GroNLP/bert-base-dutch-cased}}~\citep{de2019bertje}
and construct an approximate nearest-neighbor (ANN) index over the vocabulary.\footnote{An ANN index is a data structure designed to quickly retrieve the nearest vectors to a query vector without computing all pairwise distances, which would be prohibitively expensive in high-dimensional spaces.} 
For each token, we sample a noise vector from a $\gamma$ distribution with scale $1/\epsilon$, add it to the token’s embedding, and select the closest replacement from the ANN index. 


\paragraph{RANTEXT}
RANTEXT~\citep{tong2025inferdptprivacypreservinginferenceblackbox} employs an LLM to generate embedding vectors for tokens (we use the Dutch GPT-2\footnote{\url{https://huggingface.co/GroNLP/gpt2-small-dutch}} from~\citealt{devries2020good}).  For each token, a Laplace distribution is used to dynamically determine the size of a random adjacency list, and then new tokens are sampled from this list to replace the original token. Candidate replacements within the noise radius are selected via the exponential mechanism, favoring tokens closer to the original in embedding space. 

\subsection{De-identification pipelines}
\label{sec:method-pipelines}

We investigate five different \textit{pipelines}, i.e., automatic methods to de-identify the text in the raw dataset (\datasetraw{}).
Our dataset, as detailed in \S\ref{sec:method-dataset}, is very small; thus, all pipelines we propose next require models without fine-tuning. 
The pipelines are:
\noindent\textbf{\datasetrawNER{}}: Apply NER de-identification to \datasetraw{}.
\noindent\textbf{\datasetLLM{}}: Apply LLM de-identification to \datasetraw{}.
\noindent\textbf{\datasetrawdp{}}: Apply DP-based de-identification to \datasetraw{}.
\noindent\textbf{\datasetNERdp{}}: Apply a NER model to \datasetraw{} and apply DP-based de-identification to its output.
\noindent\textbf{\datasetLLMdp{}}: Apply LLM de-identification to \datasetraw{} and apply DP-based de-identification to its output.

In all experiments with pipeline \textbf{\datasetLLMdp{}}, we use Deepseek-70B as it is the best-performing open-source LLM in terms of privacy.


\subsection{Privacy Leakage Evaluation}\label{sec:intrinsic_evaluation}
We quantify the privacy leakage of a de-identification pipeline as \textit{the percentage of personally identifiable information (PII) leaked in the de-identified text compared to \datasetmanual{}}.
PIIs are not all the same, and the leakage of different types of PII can, in practice, have very different impact in terms of privacy preservation. 
For that reason, in this work we differentiate between leakage of \textit{direct} and \textit{indirect PII}.

\paragraph{Direct vs. Indirect PII}
Direct PII (e.g., a patient's name or telephone number) can directly identify an individual. For that reason, their leakage is the most damaging in terms of privacy preservation.
Indirect PII have lower risk but can also be used to identify an individual, usually in combination with other indirect PII or with extra auxiliary information, e.g., \citet{sweeney2000simple}'s classical example whereby zip code, date of birth
and gender is reported to uniquely identify 87\% of the population in the USA.

According to~\citet{hipaa}, the following identifier types are classified as direct PII:
name;
address (all geographic subdivisions smaller than state, including street address, city county, and zip code); 
all date-related  elements (except years) related to an individual (including birth date, admission date, discharge date, date of death, and exact age if over 89);
telephone number;
fax number;
email address;
social security number;
medical record number;
health plan beneficiary number;
account number;
certificate or license number;
vehicle identifiers and serial numbers, including license plate numbers;
device identifiers and serial numbers;
web URL;
internet protocol (IP) address;
finger or voice print;
photographic image;
and any other characteristic that could uniquely identify the individual.

In our data, we consider name of doctor, electronic health record ID, date of birth, name of patient, city, and telephone number as direct PIIs, and all the rest in Table~\ref{tab:placeholders} as indirect PIIs.

In all our experiments, we apply a strict evaluation setting where the leakage of any subword token part of a PII entity span is treated as leakage.

\subsection{Utility Evaluation}\label{sec:extrinsic_evaluation}
We assess the preservation of relevant clinical information by using the generated notes in two downstream tasks: classifying entity spans in a clinical text as drug or disorder mentions, and predicting whether a drug--disorder entity pair denotes an adverse drug event (ADE).\footnote{We note that originally the Dutch ADE dataset was collected to support these two tasks~\citep{murphy2025detection}, making these tasks ideal for utility evaluation.}
We use the original train/dev/test splits in the Dutch ADE corpus for training, model selection, and testing, respectively. For configuration details please see Appendix~\ref{appendix:utility-config}
We use the annotated entities and relations that remain in the generated texts for training and development, and evaluate the model against the original test set with gold annotations.
At low $\epsilon$ values under some settings, i.e. $\epsilon$$\leq$128 for metric-DP and $\epsilon$$\leq$16 for RANTEXT, we only have less than 10 annotated relations in the generated text, making it impossible to train the relation classification model. Thus, we omit these experiments and report performance using macro F-1 score for both entity and relation classification.
We do not report new confidence intervals for the downstream tasks because its protocol and uncertainty were established in prior work~\citep{murphy2025detection}, which already provides CIs under the same evaluation procedure.
\section{Results}

\begin{figure}[t]
     \centering
     \includegraphics[width=\linewidth]{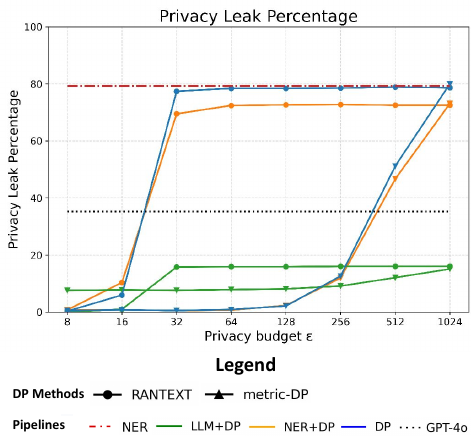}
     \caption{Comparison of privacy leakage across different de-identification pipelines and DP budgets ($\epsilon$). This figure includes two DP mechanisms: RANTEXT and Metric-DP, each applied to three pipelines: \datasetrawdp{}, \datasetNERdp{}, and \datasetLLMdp{}. For \datasetLLMdp{}, we use Deepseek-70B as the de-identification module as it performs the best in terms of privacy. Horizontal lines indicate non-DP baselines, including one NER-based pipeline (GLiNER) and one best performing LLM (GPT-4o). As $\epsilon$ increases, privacy leakage of applying DP directly to raw text also significantly increases. Meanwhile, combining DP with 
     LLM-based redaction largely enhances privacy even with high $\epsilon$ values.}
     \label{fig:privacy-epsilon}
\end{figure}

\begin{figure*}[t]
     \centering
     \includegraphics[width=\linewidth]{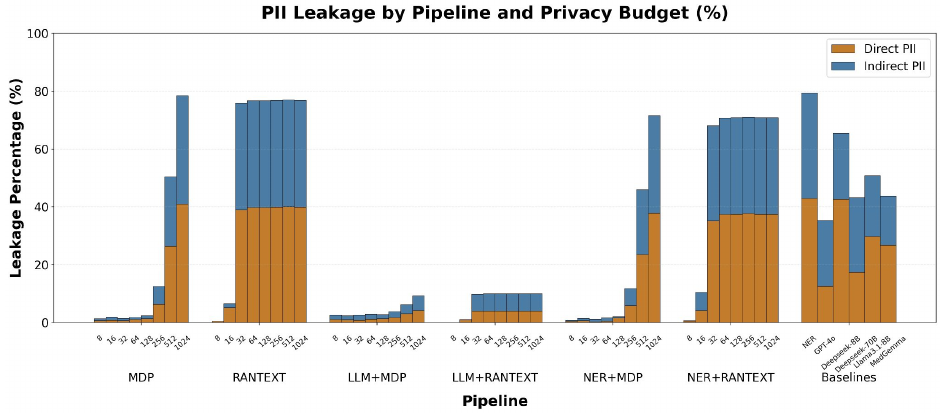}
     \caption{PII leakage by pipeline and privacy budget $\epsilon$. Bars are stacked (dark: direct PII; light: indirect PII). Across budgets, metric-DP (MDP) leaks substantially less than RANTEXT, and all methods show increasing leakage as $\epsilon$ grows, with the steepest rise for RANTEXT variants. Prepending an LLM rewrite reduces leakage for both families, with LLM+MDP achieving the lowest overall rates, followed by LLM+RANTEXT, while plain RANTEXT and NER+RANTEXT exhibit high direct leakage. NER and LLMs alone without DP also remain markedly leakier.}
     \label{fig:pii_type_leakage}
\end{figure*}

\begin{figure}[t]
     \centering
     \includegraphics[width=\linewidth]{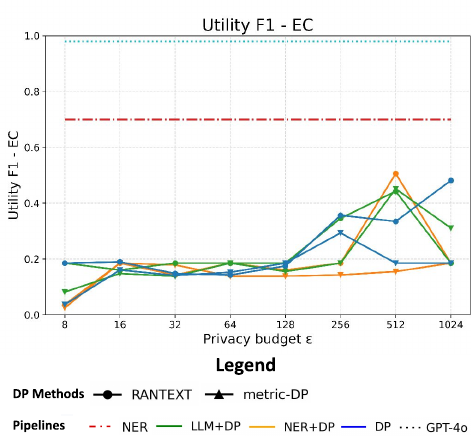}
     \caption{Comparison of utility F1-score for Entity Classification (EC) task across different de-identification pipelines and DP budgets ($\epsilon$) (see Figure~\ref{fig:privacy-epsilon} for more details). It shows that the utility scores improve with higher $\epsilon$, but still cannot recover baseline performances.}
     \label{fig:ec-epsilon}
\end{figure}


We now report on our experiments regarding the empirical privacy leakage of PII in the generated clinical notes (\ref{sec:results_privacy_leakage}), and the downstream evaluations according to drug and disorder (entity) classification  (\ref{sec:results_utility_entity}) and adverse drug event (relation) classification  (\ref{sec:results_utility_relation}).

\subsection{Privacy Leakage Results}\label{sec:results_privacy_leakage}

Figure~\ref{fig:privacy-epsilon} presents privacy leakage percentage for various pipeline configurations across DP budgets ($\epsilon$), spanning from 8 to 1024. We acknowledge that with $\epsilon$ higher than 10, the theoretical guarantees of DP fade away, but it is also usual in practical setting to work with high epsilons in order to get meaningful empirical results, as seen in works like DP-BART \citep{igamberdiev-habernal-2023-dp}. As $\epsilon$ increases, privacy leakage also rises, demonstrating the trade-off between reduced noise (higher $\epsilon$) and weaker privacy preservation. At high privacy budgets, the protection provided by DP diminishes, leading to increased privacy leakage across all pipelines.
Among the DP-based pipelines, \datasetrawdp{} consistently exhibits the highest privacy leakage across all $\epsilon$ values.
In contrast, pipelines that combine DP with NER preserve privacy for slightly higher privacy budgets, and LLM-based de-identification show significantly lower empirical leakage. Notably, pipelines with LLM preprocessing significantly outperform other approaches with a high budget. This improvement can be attributed to the strong ability of LLMs to identify and redact sensitive information more effectively, reducing the amount of private data exposed to DP noise.

In our next experiments, 
we use the Health Insurance Portability and Accountability Act~\citep[HIPAA;][]{garfinkel2015identification} to differentiate between leakage of \textit{direct} (i.e., exact identifiers such as names, emails, phone numbers) versus \textit{indirect PII} (i.e., \textit{quasi-identifiers} and other attributes that enable re-identification when linked). Figure~\ref{fig:pii_type_leakage} illustrates the leakage rates of direct and indirect PIIs for each privacy budget $\epsilon$ across pipelines.
Overall, leakage increases with $\epsilon$, but the extent and composition differ largely according to the pipeline. Metric-DP consistently leaks far less PII than RANTEXT at comparable budgets: while RANTEXT and NER+RANTEXT reach ~38–40\% direct (and ~76–71\% total) leakage at high $\epsilon$, metric-DP (MDP) better preserves privacy more at low-to–mid $\epsilon$, and only approaches RANTEX when $\epsilon$ is large. Moreover, leakage rises with $\epsilon$ for all methods, but the slope differs. RANTEXT variants escalate quickly, with the direct component dominating the increase, whereas MDP grows more gradually until the largest budgets, where its bars steepen.

Moreover, the baseline results of LLMs and NER confirms that stand‑alone methods without DP are markedly leakier: the pure NER system shows high overall leakage with a large direct component, and LLM‑only variants also leak substantially. 
However, inserting an LLM de-identification before the DP step shows a clear advantage for both families. 
Finally, pipelines combining LLMs with DP (LLM+MDP and LLM+RANTEXT) were the only pipelines we investigated that managed to keep empirical privacy leakage below 10\%, the majority of which consisted of indirect PII.

\subsection{Utility Preservation Results - Entity Classification}\label{sec:results_utility_entity}

Figure~\ref{fig:ec-epsilon} illustrates the relationship between privacy budgets ($\epsilon$) and entity classification F1 scores (utility) across different pipelines. At very low privacy budgets ($\epsilon$ $\leq$ 32), all DP pipelines demonstrate poor utility due to the high levels of noise, but utility improves as $\epsilon$ increases, peaking around $\epsilon$ = 512. \datasetrawdp{} consistently exhibits lower F1 scores across all privacy budgets, demonstrating utility degradation caused by noise addition, particularly at low $\epsilon$ values. Introducing linguistic preprocessing improves utility, with \datasetNERdp{} achieving higher F1 scores than \datasetrawdp{}, especially at mid-to-high privacy budgets ($\epsilon$ $\geq$ 128). \datasetLLMdp{} further outperforms \datasetNERdp{} in most cases, showcasing the advantage of leveraging advanced language models for preprocessing. Both DP methods (RANTEXT and metric-DP) exhibit similar trends within each pipeline, suggesting that preprocessing, rather than the DP mechanism itself, plays a more critical role in improving utility. The LLM baseline maintains consistently high F1 scores near 1.0, thereby serving as an upper bound for utility.

\subsection{Utility Preservation Results - Relation Classification}\label{sec:results_utility_relation}

\begin{figure}[t]
     \centering
     \includegraphics[width=\linewidth]{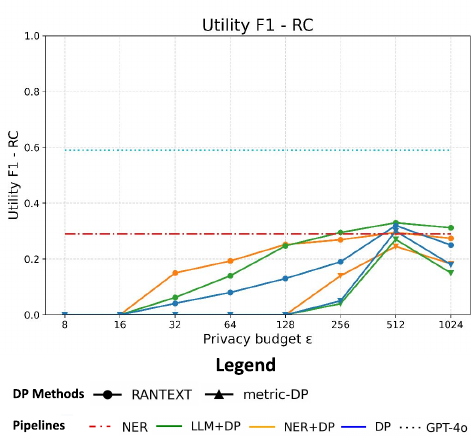}
     \caption{Comparison of utility F1-score for Relation Classification (RC) task across different de-identification pipelines and DP budgets ($\epsilon$) (see Figure~\ref{fig:privacy-epsilon} for more details). It shows that the utility scores slightly improve with higher $\epsilon$, but are significantly harmed even with high $\epsilon$, and cannot recover baseline performances.}
     \label{fig:rc-epsilon}
\end{figure}


Figure~\ref{fig:rc-epsilon} shows relation classification F1 scores (utility) across various privacy budgets ($\epsilon$) and for different pipelines. At low privacy budgets, such as $\epsilon$ = 8 or in some cases 16, no results are obtained due to the severe obfuscation which reduces the amount of usable training data to negligible levels. As $\epsilon$ increases, utility scores gradually improve across all pipelines. \datasetrawdp{} exhibit the lowest F1 scores throughout the range of privacy budgets. In contrast, both \datasetNERdp{} and \datasetLLMdp{} achieve higher F1 scores, with \datasetLLMdp{} consistently outperforming \datasetNERdp{}, particularly at mid-to-high privacy budgets ($\epsilon$ $\geq$ 128). Non-DP baselines still maintain consistently high F1 scores and establish the upper bounds for utility. 
These results underscore the crucial role of linguistic preprocessing, particularly with LLMs, in mitigating the utility degradation inherent to DP-based methods, especially in tasks requiring complex relational reasoning.

\begin{figure}[t]
     \centering
     \includegraphics[width=\linewidth]{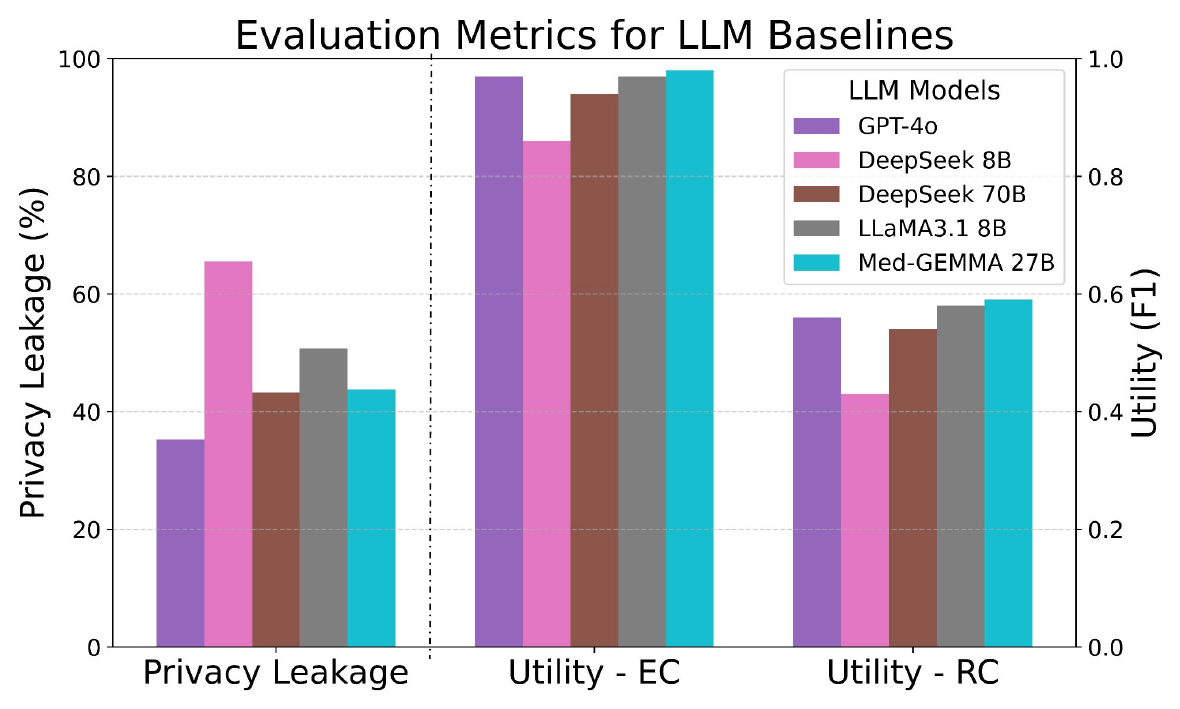}
     \caption{Comparison of evaluation metrics including Privacy Leakage, Utility - EC, and Utility - RC for \datasetLLM{} using 5 different LLMs. Privacy Leakage, where lower is better, shows that GPT-4o achieves the lowest leakage. Utility metrics for EC and RC, where higher is better, reveal that Med-GEMMA 27B consistently outperforms others while obtaining relatively low privacy leakage. GPT-4o achieves similar utility even with the lowest privacy leakage, indicating its high overall performance.}
     \label{fig:llm-barplot}
\end{figure}

\subsection{LLM Baseline Results}
Figure~\ref{fig:llm-barplot} shows the performances of privacy leakage and utility F1-scores for both EC and RC tasks among \datasetLLM{} using 5 various LLMs.
Amongst the LLM baselines, GPT-4o achieves the best, i.e. lowest privacy leakage percentage while still offering near-optimal utility for both EC and RC tasks. Med-GEMMA 27B exhibits the highest utility for both tasks while keeping the privacy leakage at a relatively low level.
This aligns with the assumption that LLMs' advanced linguistic capabilities allow them to effectively redact sensitive information.




\subsection{Operational Efficiency}

Building the manually de-identified corpus (\datasetmanual{}) was a labor-intensive process, requiring over 47 hours of annotation by two annotators~\citep{murphy2025creation}. In contrast, our automated pipelines are way more time- and cost-efficient, with metric-DP taking 3 minutes and 16 seconds per run and RANTEXT 26 minutes and 40 seconds per run.
While manual de-identification remains the gold standard for accuracy and consistency, these results highlight a substantial operational advantage for the automated methods, enabling rapid iteration over various configurations and large-scale or repeated de-identification in practice.
\section{Discussion and Conclusions}
\label{sec:conclusions}

\paragraph{General results.}
This work provides a comparative analysis of privacy-preserving de-identification for Dutch clinical notes by contrasting text-level differential privacy (DP) mechanisms, NER-based masking, and LLM-based masking, and by evaluating both privacy leakage and downstream utility on entity and relation classification tasks. Our results show a consistent tension for DP perturbation: increasing the privacy budget $\epsilon$ improves utility but increases residual leakage, and applying DP directly to raw notes yields the weakest privacy outcomes among DP pipelines. By contrast, preprocessing that removes explicit identifiers before DP (via NER or LLM masking) improves the overall trade-off, indicating that, in our no-training setting, most gains come from reducing sensitive content \emph{before} perturbation rather than relying on perturbation alone. In utility terms, token-level perturbation is especially harmful for relation classification (more than for entity classification), plausibly because relations depend on longer-range coherence and on the availability of sufficient high-quality labeled instances: under tighter privacy budgets, the perturbation can effectively corrupt the signal beyond recovery, whereas higher $\epsilon$ only partially closes the gap with non-private baselines. Finally, our operational analysis highlights the substantial gap between the cost of producing benchmark-quality manually de-identified reference data and the speed of automated pipelines, which enables rapid iteration over configurations in practice.


\paragraph{Advice for practitioners.}
When the goal is to share Dutch clinical narratives while explicitly controlling privacy risk, our results suggest prioritizing hybrid strategies: use a strong high-recall masking stage (LLM-based when feasible, otherwise NER-based) and then apply DP as a second-stage safeguard on the residual text, rather than applying DP directly to raw notes. This design reduces the sensitive surface exposed to DP and consistently improves both leakage and downstream utility relative to DP-on-raw-text, while retaining an explicit privacy knob through $\epsilon$. Practitioners should also anticipate that preserving utility for relation-centric tasks will be harder than for entity-centric tasks under text perturbation, and should validate performance on the intended downstream use case (not only on intrinsic leakage). Finally, even when automated pipelines are adopted, a manually de-identified reference (or equivalent audit protocol) remains important for trustworthy evaluation and monitoring. Although the best-performing LLM in our experiments is proprietary (GPT-4o), recent work on Italian clinical notes has shown that smaller, open-source LLMs can achieve competitive de-identification performance in a similar zero-shot setting \citep{miranda-etal-2025-mamma}, suggesting that our hybrid pipeline design is not inherently tied to proprietary models.

\paragraph{Future work.}
Three extensions are particularly important: (i) expanding beyond Dutch to assess whether the same trade-offs hold across languages with different resource availability; (ii) enabling broader experimentation and external validation via public benchmarks or carefully designed synthetic alternatives; and (iii) widening utility evaluation beyond entity/relation classification to additional clinical NLP use cases (e.g., document-level prediction), where sensitivity to perturbation may differ. More generally, future work should also strengthen adversarial evaluation of re-identification risk and explore hybrid designs that better target quasi-identifiers without unnecessarily corrupting clinical signals.

\section*{Limitations}

While our study provides valuable insights into the application of differential privacy (DP) for clinical text de-identification, it has certain limitations. First, our work focuses exclusively on Dutch clinical narratives, which may limit the generalizability of our findings to other languages or multilingual clinical datasets. Expanding this research to additional languages would provide a broader understanding of the effectiveness and scalability of the proposed techniques. Second, the dataset used in this study is private, as it contains sensitive medical information, which restricts reproducibility and external validation by other researchers. Although this is a common constraint in clinical research, the use of publicly available or synthetic benchmark datasets in future studies could help foster more extensive experimentation and comparison. Third, our privacy leakage metric - residual PII percentage against a manually de-identified reference - is an empirical proxy, not a formal privacy measure. While DP provides theoretical guarantees on output indistinguishability, empirical leakage captures a complementary quantity: how much identifiable information survives in practice. These two notions should not be conflated, yet empirical evaluation against a gold-standard reference remains one of the few viable approaches for assessing de-identification effectiveness on real data. Lastly, while we evaluated our pipelines on two downstream tasks (entity and relation classification), further exploration of their performance across a wider variety of use cases, such as document classification or clinical outcome prediction, would provide a more comprehensive assessment of their utility.

\section{Bibliographical References}\label{sec:reference}

\bibliographystyle{lrec2026-natbib}
\bibliography{custom,anthology}

\label{lr:ref}
\bibliographystylelanguageresource{lrec2026-natbib}
\bibliographylanguageresource{languageresource}

\appendix
\section{Appendix}

\subsection{LLM De-identification Prompts and Performances}
\label{appendix:llm-prompt}

We use the following prompt for using LLMs as de-identification tools: "\textit{Please anonymize the following clinical note. Replace all the following information with corresponding masks: Replace hospital names with <ZIEKENHUIS>; Replace abbreviations for hospitals (e.g. ZKH) with <ZKH>; Replace doctor names with <ARTS>; Replace city names with <STAD>; Replace signs or signals (SEIN) with <SEIN>; Replace patient names with <NAAM>; Replace entries from electronic health records with <EHR>; Replace rare diseases with <RARE\_DISEASE>; Replace clinical trial identifiers with <TRIAL-ID>; Replace holidays with <FEESTDAG>; Replace rare disease treatments with <RARE\_DISEASE\_TREATMENT>; Replace department names with <AFDELING>; Replace rehabilitation centers with <REVALIDATIECENTRUM>; Replace sickbay references with <ZIEKENBOEG>; Replace telephone numbers with <TELNR>; Replace pharmacies or drug stores with <APOTHEEK>; Replace dates of birth with <GEBOORTEDATUM>.}"

Table~\ref{tab:llm-comparison} provides a comparison of macro F1-scores, true positives (TP), false positives (FP), and false negatives (FN) across various LLM models applied to the redaction task. DeepSeek-8b and DeepSeek-70b exhibit relatively low F1-scores of 0.337265 and 0.337807, respectively, with DeepSeek-70b achieving slightly higher true positives (TP = 499) but at the cost of significantly more false positives (FP = 3614).
LLaMA3-8b demonstrates a notable improvement with a good balance between true positives and false positives. 
MedGemma emerges as the best-performing model, with the highest macro F1-score of 0.591672. This is supported by its significant reduction in false positives and the highest true positives.

\begin{table}[t]
\centering
\begin{tabular}{|c|c|c|c|c|}
\hline
\textbf{Model} & \textbf{F1-score} & \textbf{TP} & \textbf{FP} & \textbf{FN} \\ \hline
GPT-4o & 0.205839 & 569 & 4848 & 310 \\ \hline
DeepSeek-8b & 0.337265 & 303 & 718 & 576 \\ \hline
DeepSeek-70b & 0.337807 & 499 & 3614 & 380 \\ \hline
LLaMA3-8b & 0.492872 & 433 & 652 & 446 \\ \hline
MedGemma & 0.591672 & 495 & 399 & 384 \\ \hline
\end{tabular}
\caption{Comparison of LLMs by F1-scores, True Positives (TP), False Positives (FP), and False Negatives (FN).}
\label{tab:llm-comparison}
\end{table}

\subsection{Downstream Evaluation Configuration}
\label{appendix:utility-config}

For the downstream Entity Classification (EC) and Relation Classification (RC) tasks, we use MedRoberta.nl~\citep{verkijk2021medroberta} as the base model due to its best performance according to ~\citet{murphy2025detection} and train with the remaining entity/relation in our generated data. We use a batch size of 8, a learning rate of 3e-5, a warmup ratio of 0.2 for EC and a batch size 128, hidden sizes of 512, 128, 32, a dropout rate of 0.5, a learning rate 1e-6, and a patience of 30 for RC.
We report the micro F1-score of EC and macro F1-score of RC as annotated relations suffer from high imbalance.
Table~\ref{tab:ec-results} and Table~\ref{tab:rc-results} show the full results of the EC and RC evaluation.

\begin{table*}[htbp]
\centering
\begin{tabular}{lccccccccc}
\toprule
\textbf{Pipeline/Epsilon} & 8 & 16 & 32 & 64 & 128 & 256 & 512 & 1024 & $\infty$ \\ \hline
Baseline & -- & -- & -- & -- & -- & -- & -- & -- & 0.91 \\ \hline
NER & -- & -- & -- & -- & -- & -- & -- & -- & 0.70 \\ \hline
LLM & -- & -- & -- & -- & -- & -- & -- & -- & 0.97 \\ \hline
metricDP & 0.0368 & 0.1605 & 0.1412 & 0.1524 & 0.1854 & 0.2933 & 0.1854 & 0.1854 & -- \\ \hline
RANTEXT & 0.1854 & 0.1898 & 0.1479 & 0.1427 & 0.1751 & 0.3567 & 0.3348 & 0.4812 & -- \\ \hline
NER+metricDP & 0.0249 & 0.1855 & 0.1787 & 0.1383 & 0.1388 & 0.1427 & 0.1552 & 0.1854 & -- \\ \hline
NER+RANTEXT & 0.0368 & 0.1854 & 0.1427 & 0.1854 & 0.1605 & 0.1854 & 0.5061 & 0.1854 & -- \\ \hline
LLM+metricDP & 0.0817 & 0.1479 & 0.1383 & 0.1854 & 0.1552 & 0.1854 & 0.4514 & 0.3110 & -- \\ \hline
LLM+RANTEXT & 0.1854 & 0.1605 & 0.1854 & 0.1854 & 0.1854 & 0.3457 & 0.4426 & 0.1854 & -- \\
\bottomrule
\end{tabular}
\caption{Performance of Entity Classification task across privacy budgets ($\epsilon$). Dashes indicate settings where $\epsilon$ is not applicable.}
\label{tab:ec-results}
\end{table*}

\begin{table*}[htbp]
\centering
\begin{tabular}{lccccccccc}
\toprule
\textbf{Pipeline/Epsilon} & 8 & 16 & 32 & 64 & 128 & 256 & 512 & 1024 & $\infty$ \\ \hline
Basline & -- & -- & -- & -- & -- & -- & -- & -- & 0.62 \\ \hline
NER & -- & -- & -- & -- & -- & -- & -- & -- & 0.29 \\ \hline
LLM & -- & -- & -- & -- & -- & -- & -- & -- & 0.59 \\ \hline
metricDP & -- & -- & -- & -- & -- & 0.050 & 0.300 & 0.180 & -- \\ \hline
RANTEXT & -- & -- & 0.041 & 0.080 & 0.130 & 0.190 & 0.320 & 0.250 & -- \\ \hline
NER+metricDP & -- & -- & -- & -- & -- & 0.140 & 0.245 & 0.183 & -- \\ \hline
NER+RANTEXT & -- & -- & 0.150 & 0.193 & 0.252 & 0.269 & 0.295 & 0.274 & -- \\ \hline
LLM+metricDP & -- & -- & -- & -- & -- & 0.040 & 0.270 & 0.150 & -- \\ \hline
LLM+RANTEXT & -- & -- & 0.062 & 0.140 & 0.247 & 0.295 & 0.330 & 0.312 & -- \\
\bottomrule
\end{tabular}
\caption{Performance of Relation Classification task across privacy budgets ($\epsilon$). Dashes indicate settings where $\epsilon$ is not applicable.}
\label{tab:rc-results}
\end{table*}

\end{document}